\documentclass[preprint,12pt]{elsarticle}

 \usepackage{graphics}
 \usepackage{graphicx}
 \usepackage{epsfig}
\usepackage{float}
\usepackage{subfloat}
\usepackage{caption}
\usepackage{subcaption}
\usepackage{amsmath,amsfonts,amssymb}

\usepackage{amssymb}

\usepackage{ulem} 
\usepackage{lineno}
\usepackage{xcolor}

\usepackage{bm} 
\newcommand{\nb}{\nobreak\hspace{.03333em plus .016667em}}
\newcommand{\nbl}{\nobreak\hspace{.1em plus .05em}}

\journal{Materials Science \& Engineering A}

\begin{document}

\begin{frontmatter}



\title{
The analysis of internal stresses on a welded joint in Grade 91 steel under creep test loading: 
synchrotron X-ray diffraction measurements and modeling
}



\author[a1]{Solenne Collomb}
\author[a2,a3,a4]{Xiaolei Chen}
\author[a1]{Jean-Philippe Tinnes}
\author[a1,a3]{Thomas Schenk \corref{cor1}}
\ead{thomas.schenk@univ-lorraine.fr}
\author[a1]{Olivier Ferry}
\author[a1]{Tsareva Svetlana}
\author[a1]{Abdelkrim Redjaïmia}
\author[a1,a3]{Alain Jacques$^{\nb\dagger}$}

\address[a1]{Université de Lorraine, CNRS, IJL, F-54000 Nancy, France}
\address[a2]{LEM3 Laboratory, Université de Lorraine, CNRS, Arts et Métiers ParisTech, F-57000 Metz, France}
\address[a3]{LABoratoire d'EXcellence DAMAS, Université de Lorraine, F-57000, Metz, France}
\address[a4]{Université Paris-Saclay, CentraleSupélec, ENS Paris-Saclay, CNRS, LMPS, Gif-sur-Yvette, France}
%

\cortext[cor1]{Corresponding author}

\begin{abstract}
The analysis and understanding of creep damage of Grade 91 steel welded joints is an important topic 
in the energy industry. 
Creep tests on welded joints were carried out at 600\nb$^{\circ}$C, 100\nbl{MPa} and then interrupted at 0\%, 10\%, 30\%, 50\%, 80\%
of the expected life and after failure. 
Creep damage is characterised by cavity bands located exclusively in the core of the sample in the InterCritical Heat Affected Zone (ICHAZ). 
These samples were tested using \textit{in situ} synchrotron XRD along the welded joint under creep conditions for the different creep life time.
The experimental results show a significant strain evolution and creep damage characteristic 
on the welded joint, with a local maximum at the Heat Affected Zone (HAZ).
Following this, a finite element creep strain analysis was performed for comparison with the experimental results. 
\end{abstract}


\end{frontmatter}


\section{Introduction}\label{S:1}
Martensitic steels such as Grade 91, 9Cr1Mo-NbV, are used to build components of thermal power plants.
They indeed exhibit exceptional creep properties between 540\nb$^{\circ}$C and 610\nb$^{\circ}$C~\cite{ennis_recent_2003},
after a fine-tuned thermal treatment~\cite{abd_el-azim_long_2013}. Welding is commonly used to join these Grade 91 power-plant components.
However, this method leads to significant changes of their mechanical properties and also to an unusual damage mode near the welded joint. At high temperature and low stress, damage is present with cavities and a Type IV region fracture, localized in the Heat Affected Zone (HAZ), close to the Base Metal (BM), \textit{i.e.} the InterCritical Heat Affected Zone (ICHAZ)~\cite{watanabe_creep_2006, li_evaluation_2009}. The ICHAZ microstructure is composed of coarse previous martensitic grains 
and new fine martensitic grains as well as a wide range of precipitates with different sizes.
This heterogeneous zone is comparatively weaker than the other weld zones and accumulates the creep damage in the form of creep void.
Different authors observe an initiation of cavities in the HAZ at about 20\% of the lifetime. The number of cavities increases
up to about 70\% of the lifetime, then the cavities coalesce around 80\% of the lifetime to form cracks
~\cite{li_evaluation_2009, siefert_evaluation_2016}.
Some authors focussed on the internal stress distribution in order to explain this cavitation phenomenon:
the local concentration of plastic strain and the high triaxiality rate might influence the distribution of cavities
~\cite{watanabe_creep_2006, ogata_damage_2009}. The heterogeneity of the ICHAZ microstructure results in significant multiaxial stresses
~\cite{li_evaluation_2009, abd_el-azim_creep_2013}. Ogata~\cite{ogata_damage_2009} conducted internal pressure creep test on longitudinal welded
joint and reported that the higher voids number density is localised at the mid-thickness in the HAZ, where the circumferential
stress and the stress triaxiality are highest.
According to the authors, the heterogeneity in the welded joint affects the stress and strain distribution, 
which modifies the creep damage kinetics~\cite{ogata_damage_2010, andersson_significance_1998}. 
Multi-materials finite element calculations support this stress effect in the welded joint,
except that the creep properties of each constituent
have been only deduced from the BM, which are not systematically determined by experiments~\cite{gaffard_modelling_2008}.
Thus, it is necessary to determine the physical mechanisms responsible for the damage in each zone before modelling them
(damage controlled by principal stress or Von Mises stress).
The purpose of the present paper is thus to study the evolution of internal stresses during creep, and their influence on the initiation
of the cavitation phenomenon is further investigated. For this purpose, we performed \textit{in situ}
synchrotron XRD tests along welded joint under creep conditions at different times of their creep life.
Then a finite element creep strain analysis is conducted for comparison with the experimental results.    
 
\section{Material and Methods}\label{S:2}
\subsection{Welding process}

\begin{figure}[h]
	\centering\includegraphics[width=0.95\linewidth]{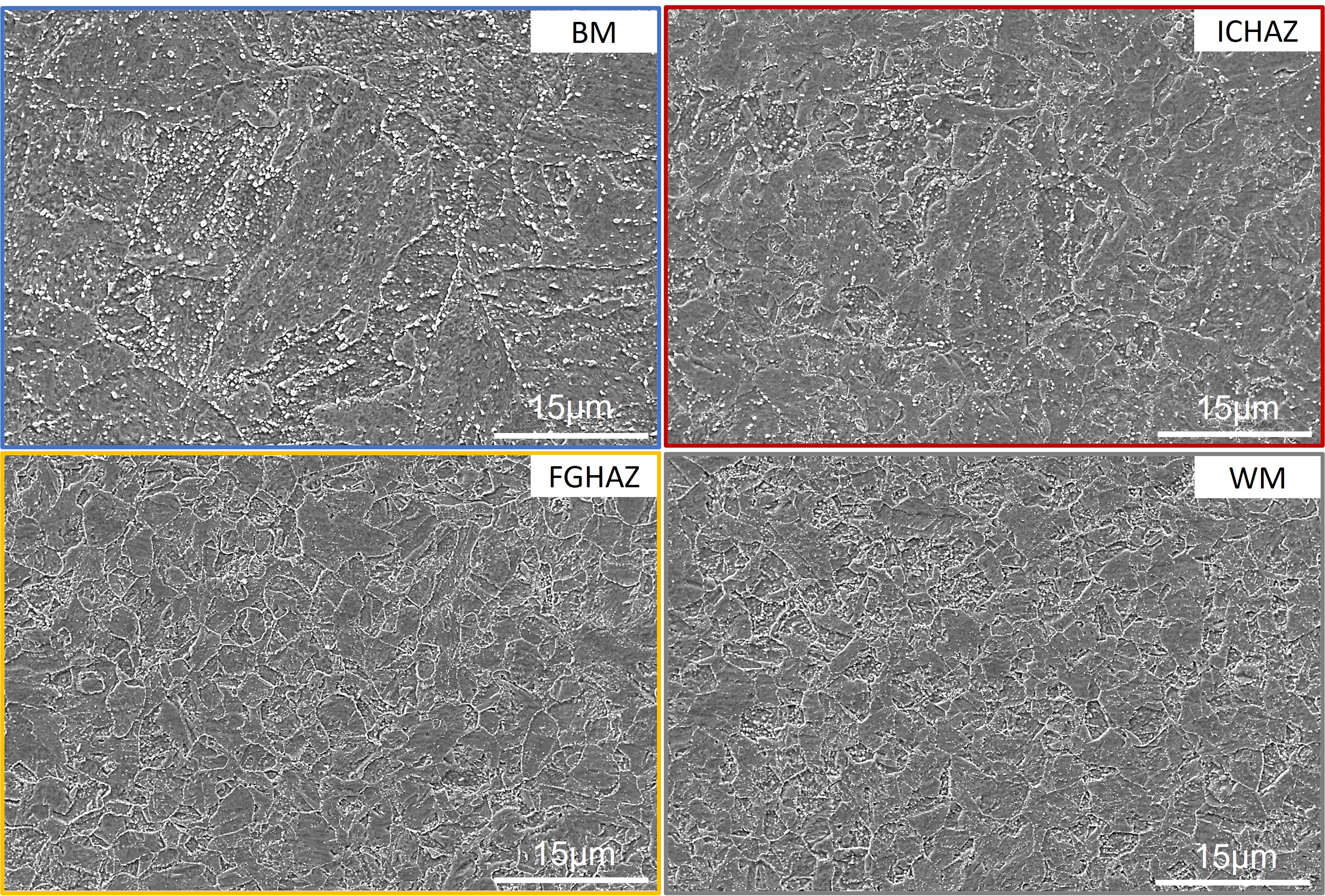}
	\caption{SEM pictures of the welded joint microstrutures for: BM, HAZ with the ICHAZ and FGHAZ, and the WM.}\label{micro}
\end{figure}

The chemical composition of ASTM Grade 91 (9Cr-1Mo-V-Nb) is shown in Table 1. The steel microstructure was obtained after an austenitization
at 1060\nb$^{\circ}$C for 60 min, air cooling, then a tempering at 785\nb$^{\circ}$C for 60 min. Chamfered 20 mm thick plate were multi-pass welded
using the Tungsten Inert Gas process (TIG), with Thermanit MTS-3 welding wire  (Table 1). The plate was pre-heated at 200\nb$^{\circ}$C before the welding
and maintained between 200\nb$^{\circ}$C and 300\nb$^{\circ}$C during the welding. A multi-layer welding was realised with 46 welding passes
at 180\nbl{A} and 8\nbl{mm/min}.
A Post-Weld Heat Treatment (PWHT) was applied at a temperature of 760\nb$^{\circ}$C during 2 hours to reduce residual stresses due to the welding process. 
There are different parts of the joint weld, \textit{i.e.}, the Weld Metal (WM),
in the center, surrounded by the HAZ, followed by the BM\@. 
The HAZ is divided in two zones, a Fine Grain HAZ (FGHAZ) and an ICHAZ,
located at the interface with the BM\@. The corresponding microstructures of these different parts observed
by Scanning Electron Microscope (SEM) are presented in Figure~\ref{micro}. Because of the multi-pass welding,
the Coarse Grain HAZ (CGHAZ) created during a pass was altered by the next pass,
transforming the microstructure into fine grains (FGHAZ). 
Cross-weld samples, with 8 mm diameter and a 50 mm gauge length were machined in the perpendicular direction of the welded joint
in the plate, to have the WM in the centre of the sample.

\begin{table}[h]
	\centering
	\begin{tabular}{l c c c c c c c c c c }
		\hline
		{Composition} & C & Si & Mn & Cr & Mo & V & Nb & Ni & Cu & N  \\
		\hline
		Base metal & 0.11 & 0.31 & 0.51 & 8.5 & 0.97 & 0.20 & 0.07 & 0.07 & 0.03 & 0.04  \\
		Weld metal & 0.10 & 0.29 & 0.5 & 9.7 & 0.93 & 0.19 & 0.06 & 0.4 & 0.03 & 0.04 \\
		\hline
	\end{tabular}
	\caption{Chemical composition of the Grade 91 and the welding wire Thermanit MTS-3 (wt\%).}
\end{table}

\subsection{Interrupted creep tests}

\begin{figure}[h]
\centering\includegraphics[width=1\linewidth]{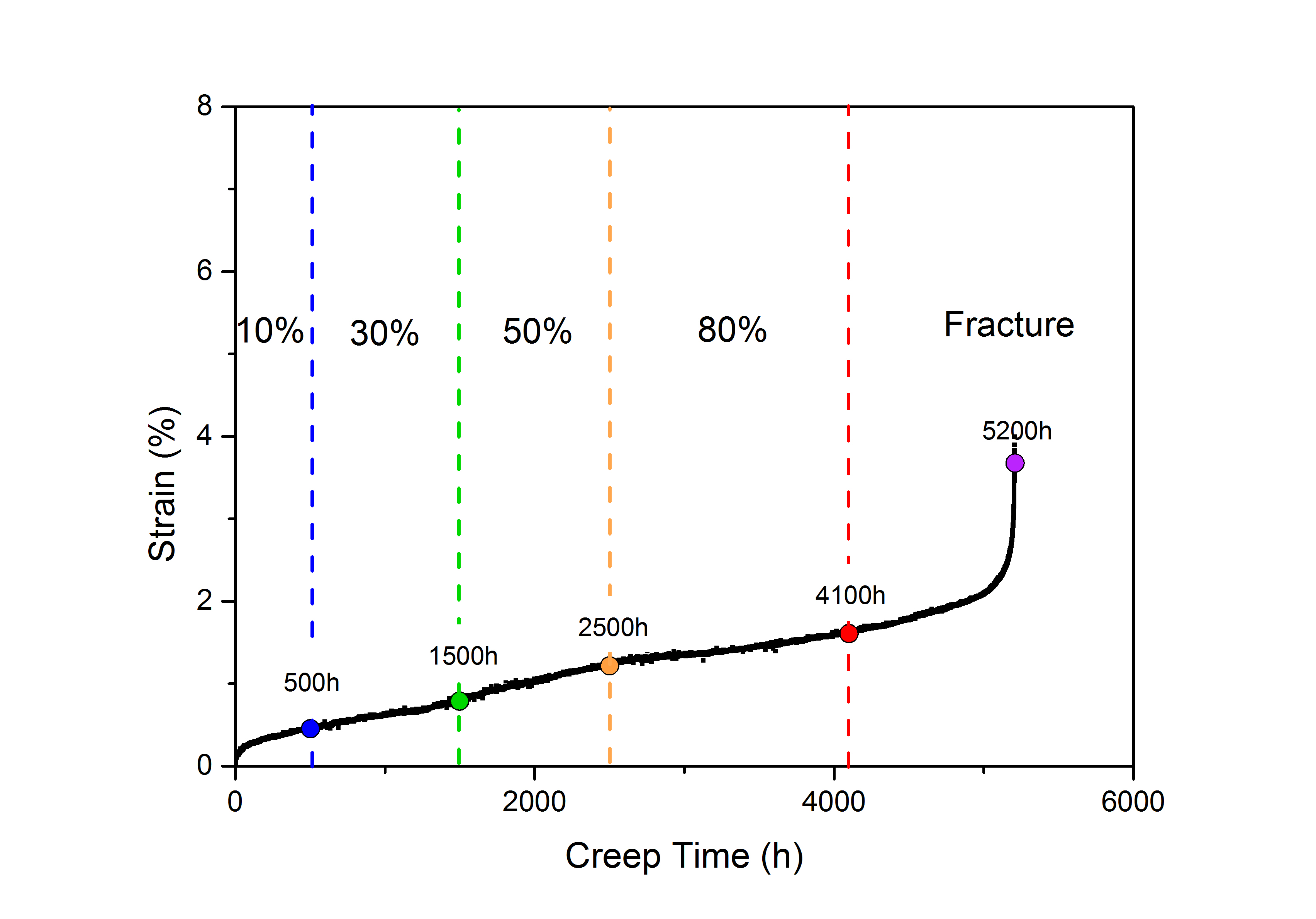}
\caption{Intermediate and failed creep tests of the welded joint sample, at 600\nb$^{\circ}$C and 100\nbl{MPa}.}\label{fig1}
\end{figure}

Creep tests were performed on welded joints according to NF EN ISO 204. The temperature was controlled with 3 thermocouples
to ensure homogeneity. The temperature gradient between the bottom and the top of the furnace did not exceed +/-2\nb$^{\circ}$C.
The crossheads displacement was measured by two extensometers, on both heads of the sample. Temperature and stress levels
have been chosen to lead to the same kind of damages as the ones observed in power plants pipes. Constant load creep tests were performed at
600\nb$^{\circ}$C, 100\nbl{MPa}, under controlled atmosphere.
Several samples were tested under the same creep conditions (temperature and stress) and the tests were interrupted at
respectively 0\%, 10\%, 30\%, 50\%, 80\% of the expected lifetime and after fracture. With these creep conditions (600\nb$^{\circ}$C and 100\nbl{MPa})
the fracture occurs after 5200 hours, as shown in Figure~\ref{fig1}.

\subsection{Synchrotron XRD tests}

\begin{figure}[h]
	\centering\includegraphics[width=1\linewidth]{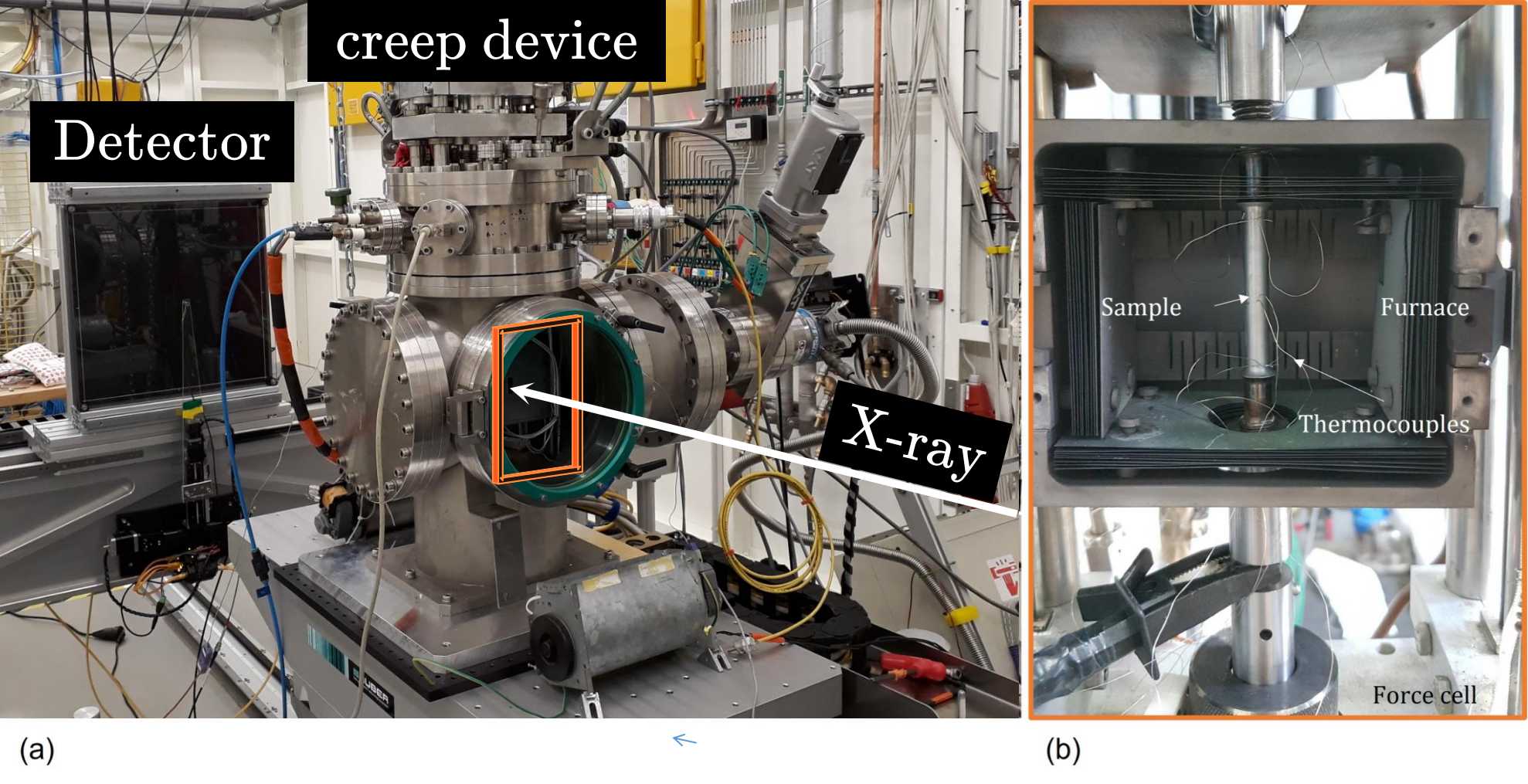}
	\caption{a) \textit{in situ} testing device on beamline P21 at the DESY PETRA III synchrotron (Germany),
	with schematic X-ray beam path, b) focus on testing device furnace,
	showing the sample with welded thermocouples.}\label{fig2}
\end{figure}

The experiments were performed on the P21.2 Swedish materials science beamline of the DESY PETRA III synchrotron
with a monochromatic beam energy of 89.59 $\pm$ 0.01 keV ($\Delta$E/E = $2\times10^{-3}$). Such a High Energy synchrotron X-Rays beamline
provides a penetrating beam which allows diffraction experiments in transmission through a 8 mm thick specimen and thus to obtain
volume information on massive samples. The diffraction rings were collected on a high-resolution 2048$\times$2048 flat panel Perkin-Elmer detector
with a 150$\times$150 $\mu m$ pixels size at a distance of 916.47 $\pm$ 0.2 mm from the specimen axis.  
To determinate internal residual stresses, the five pre-crept samples were strained at 600\nb$^{\circ}$C and 100\nbl{MPa} (creep conditions) for two hours
in a high temperature tensile device~\cite{feiereisen_mechanical_2003} transparent to X-Rays  which had been put on a heavy load diffractometer
(Figure~\ref{fig2}). Successive diffraction patterns were then recorded every 100\nbl$\mu$m along the gauge length of the specimen (50\nbl{mm}).
Two specimen orientations were used (Figure~\ref{fig3}): 0$^{\circ}$ orientation (resp. 90$^{\circ}$ orientation) with the incident beam orientation
along the $X_1$ (resp. $X_2$) axis. In the first case the incident beam travels through homogeneous areas of the specimen, while in the second one it
travels through areas with a different microstructure and stress state. 
The resulting diffracted rings are integrated along the beam path through the specimens.

\begin{figure}[h]
	\centering\includegraphics[width=1\linewidth]{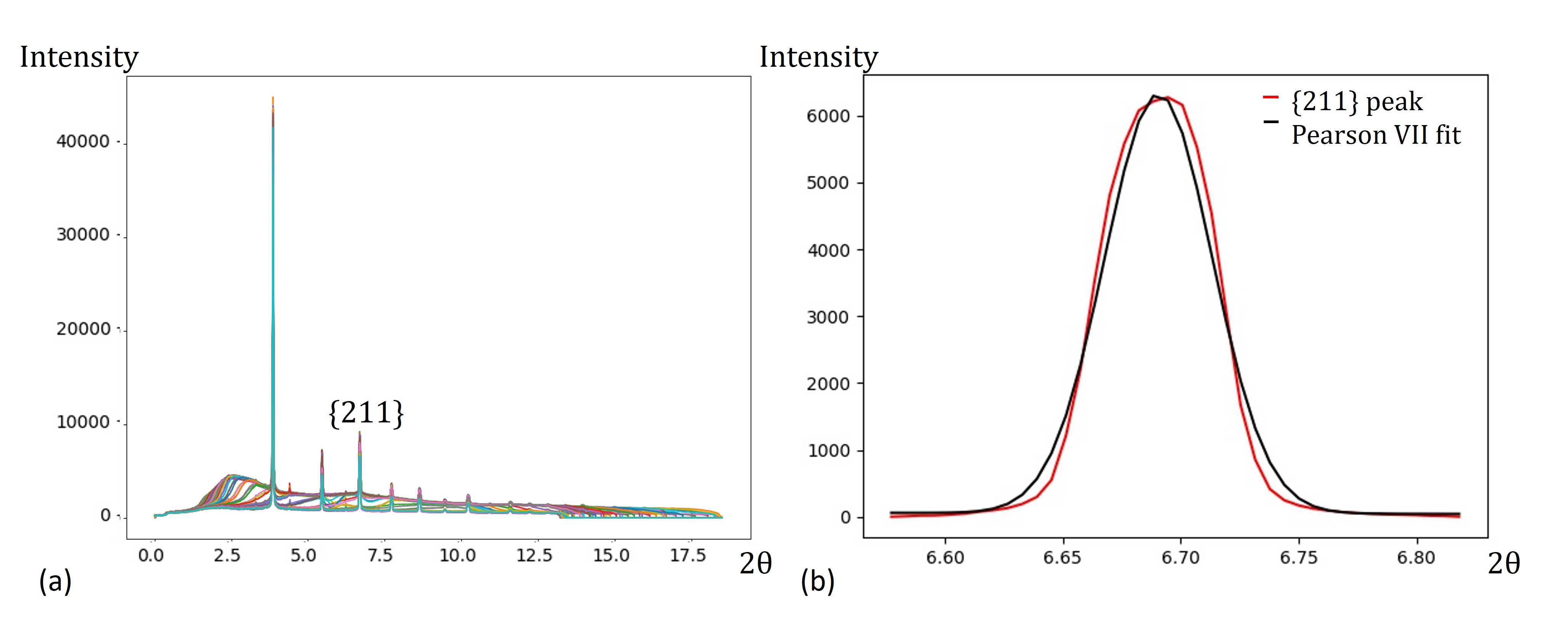}
	\caption{a) Diffractogram of intensity in function of Bragg diffraction angle $2\theta$, for one position on the welded joint
	and for the 90 sectors of the diffraction rings. b) Focus on the peak {211} with Pearson VII fit.}\label{diffracto}
\end{figure}

\begin{figure}[h]
	\centering\includegraphics[width=1\linewidth]{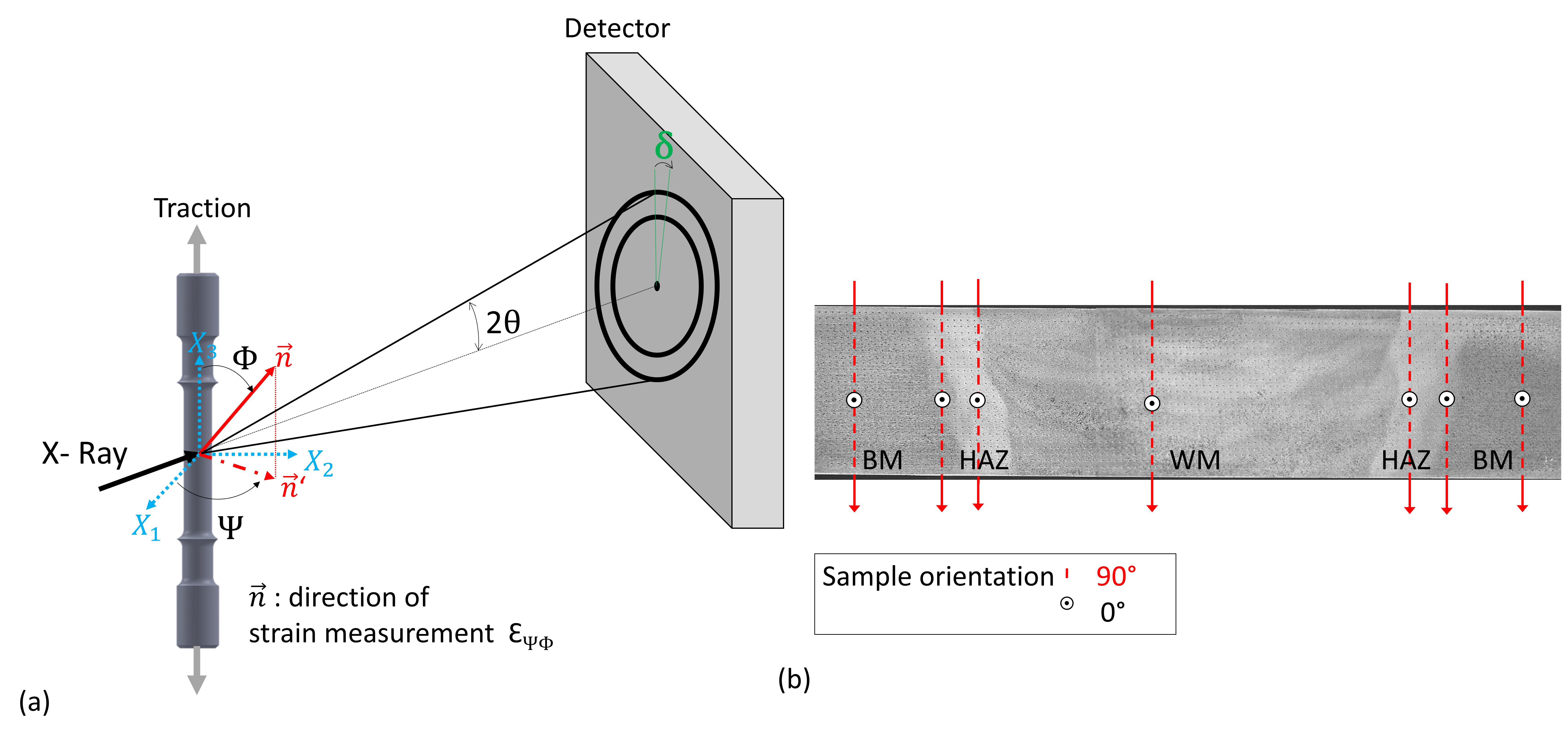}
	\caption{
		a) Illustration of the strain measurement direction and angles
		for a XRD test, b) the optical micrograph of the welded joint, 
		crossed by the X-ray beam for the 0$^{\circ}$ and 90$^{\circ}$ specimen orientation.
		}\label{fig3}
\end{figure}

Debye–Scherrer rings were corrected with dark field and calibrated, using pyFAI software~\cite{kieffer_pyfai_2013}.
The intensities were caked into 90 diffractograms of intensity in function of diffraction angle 2$\theta$, in order to precisely measure
the angular variations of the deformed diffraction peaks. The diffractogram peaks were fitted with a Pearson VII function as shown in
Figure~\ref{diffracto}. To remove uncertainties due to the variations of the beam position over time, averaging of opposite azimuths was used.

The $\sin^{2}\Psi$ method is used to determinate the internal residual stresses with the XRD experiments, 
because it is sensitive to small variations~\cite{geandier_relation_2002, he_two-dimensional_2009}. 
The angles ($\Psi$, $\Phi$, $\theta$, $\delta$) used for this method are defined in the scheme Figure~\ref{fig3} a), 
by equations~\cite{he_two-dimensional_2009, heidelbach_quantitative_1999, gelfi_x-ray_2004}:
\begin{equation}
\sin^{2} \Psi=1-(\cos\Phi^{2}\sqrt{(1-\sin^{2}\delta))}
\end{equation}
with 2$\theta$ the peak position and $\delta$ the azimuth angle on the Debye–Scherrer ring. 
In the case of triaxial stress state, the measured strain $\varepsilon_{\Psi\Phi}$ is the tensor projection
onto the measurement direction $\bm{n}$, defined by: 
\begin{align*}
\varepsilon_{\Psi\Phi}&= \bm{n} \bm{\varepsilon} \bm{n}^{t}\\
						& = \sin^{2}\Psi \cos^{2}\Phi\varepsilon_{11}+\sin^{2} \Psi \sin^{2} \Phi\varepsilon_{22}+\cos^{2}\Psi\varepsilon_{33}+\sin^{2} 
						\Psi \sin2\Phi\varepsilon_{12} \\
						& +\sin2\Psi \sin\Phi\varepsilon_{23}+\sin2\Psi \cos\Phi\varepsilon_{13}
\end{align*}
with $\varepsilon_{ij} = \frac{1+\nu}{E} \sigma_{ij} - \frac{\nu}{E} \delta_{ij} \sigma_{kk}$ and $\delta_{ij}=1 $ if $i=j$ or $0$ if $ i\ne j $.
Peak fitting was focussed on the $ \{211\} $ peak (see Figure~\ref{diffracto}b) because the radiocrystallographic elastic constants
for $ \{211\} $ peak are representative of macroscopic constants~\cite{he_two-dimensional_2009}.  

\section{Experimentals Results}\label{S:3}
\subsection{Type IV cracking}

A creep test was performed at 600\nb$^{\circ}$C and 100\nbl{MPa} until failure at 5200 hours. The sample fracture occurred in the HAZ and more especially at 
ICHAZ, close to the BM and parallel to the fusion line (see Figure~\ref{fig4}). This Type IV fracture is produced by creep cavitation. 
The fracture zone exhibits a significant necking which is also apparent in the second HAZ of the welded joint, 
with less area reduction. This necking suggests a high local viscoplastic deformation, which started during stage II of creep and increased 
during stage III\@. The major part of the specimen elongation results from this local strain.

\subsection{Cavities}
\begin{figure}[H]
	\centering\includegraphics[width=1\linewidth]{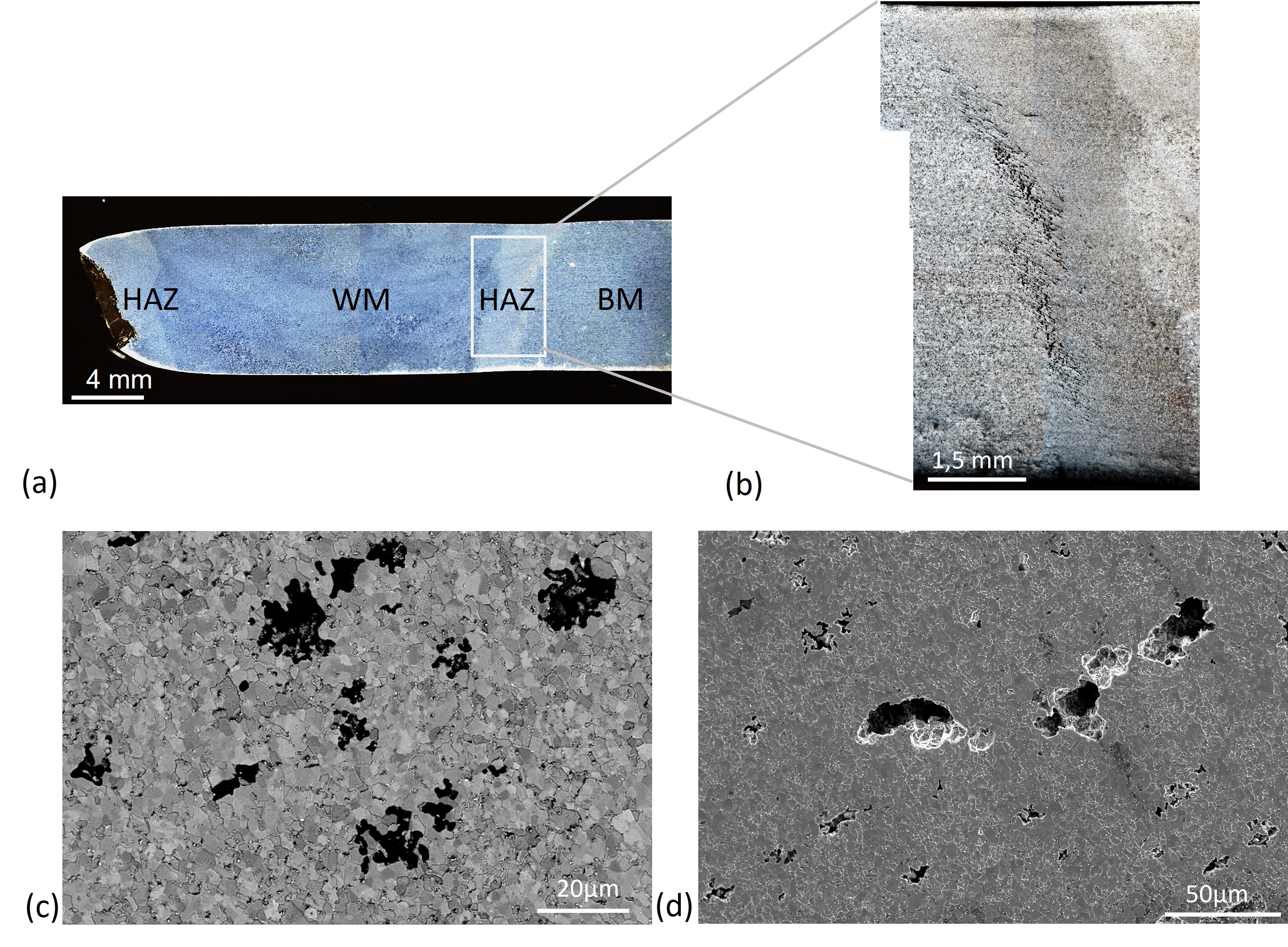}
	\caption{
		a) Optical macrograph of a cross-section along the fractured specimen length,
		after 5200 hours of creep at 600\nb$^{\circ}$C and 100\nbl{MPa}, b) zoom in the HAZ,
		c)- d) SEM pictures in the ICHAZ cavity band, in BSE and SE mode, respectively.
	}\label{fig4}
\end{figure}

The optical micrographs in Figure~\ref{fig4} show the two necking zones of the fractured sample.
A high density of cavities is observed along the fracture and in the second ICHAZ\@.
These cavity bands are located exclusively in the core of the sample
and not on the surface. 
The SEM pictures in Figure~\ref{fig4} indicate that these cavities were intergranular and appeared mainly
in the vicinity of prior austenite grain boundaries. There are three categories of cavities as pointed out by the SEM pictures: 
small isolated cavities between the grains, clusters of connected medium sized cavities and finally a coalescence of
cavities that forms the nuclei of cracks. The crack tips at grain boundaries are blunted. However, despite the high local strain, the grains
near the cavities do not seem distorted. Thus, the cavities nucleation would not be only very local visco-plastic phenomena.
These observations justify the analyses of internal stress distributions to explain the occurrence of these cavities. 
The SEM characterisation shows that cavities are clustered in a band located at 2 mm from the fusion line \textit{i.e.},
in the ICHAZ, close to the BM\@. This band is only present in the core of the sample, relatively far from the surface.
Cavitation damage starts during  stage II of creep (20-30\% of the lifetime), with the presence of a low density of small cavities.
Then, a band of cavities appears from 2500 hours (50\% of lifetime) and increases in density with creep time. The cavities size drastically
increases from 4100 hours, \textit{i.e.} during stage III creep. These results are in agreement with Li~\cite{li_evaluation_2009}.

\subsection{Internal elastic strain monitoring during creep life}
In our case, for a  $0^{\circ}~(90^{\circ})$ sample orientation the strain equation $\varepsilon_{\Psi\Phi}$ becomes: 

\begin{equation}
\begin{array}{c}
    \varepsilon_{\Psi\Phi}=\sin^{2}\Psi (\varepsilon_{11}-\varepsilon_{33})+\varepsilon_{33}+\sin2\Psi\varepsilon_{13}
\\
    (\varepsilon_{\Psi\Phi}=\sin^{2}\Psi (\varepsilon_{22}-\varepsilon_{33})+\varepsilon_{33}+\sin2\Psi\varepsilon_{23})
\end{array}
\end{equation}

The curve $\varepsilon_{\Psi\Phi}$ as a function of $\sin^{2}\Psi$ is based on the following parameters:
\begin{itemize}
\item Intercept =  $ \varepsilon_{33} $
\item Slope = $  \varepsilon_{11} - \varepsilon_{33}~~(\varepsilon_{22}- \varepsilon_{33}) $
\item Ellipse opening =$  \varepsilon_{13}~~(\varepsilon_{23}) $
\end{itemize}

\begin{figure}[H]
	\centering\includegraphics[width=0.9\linewidth]{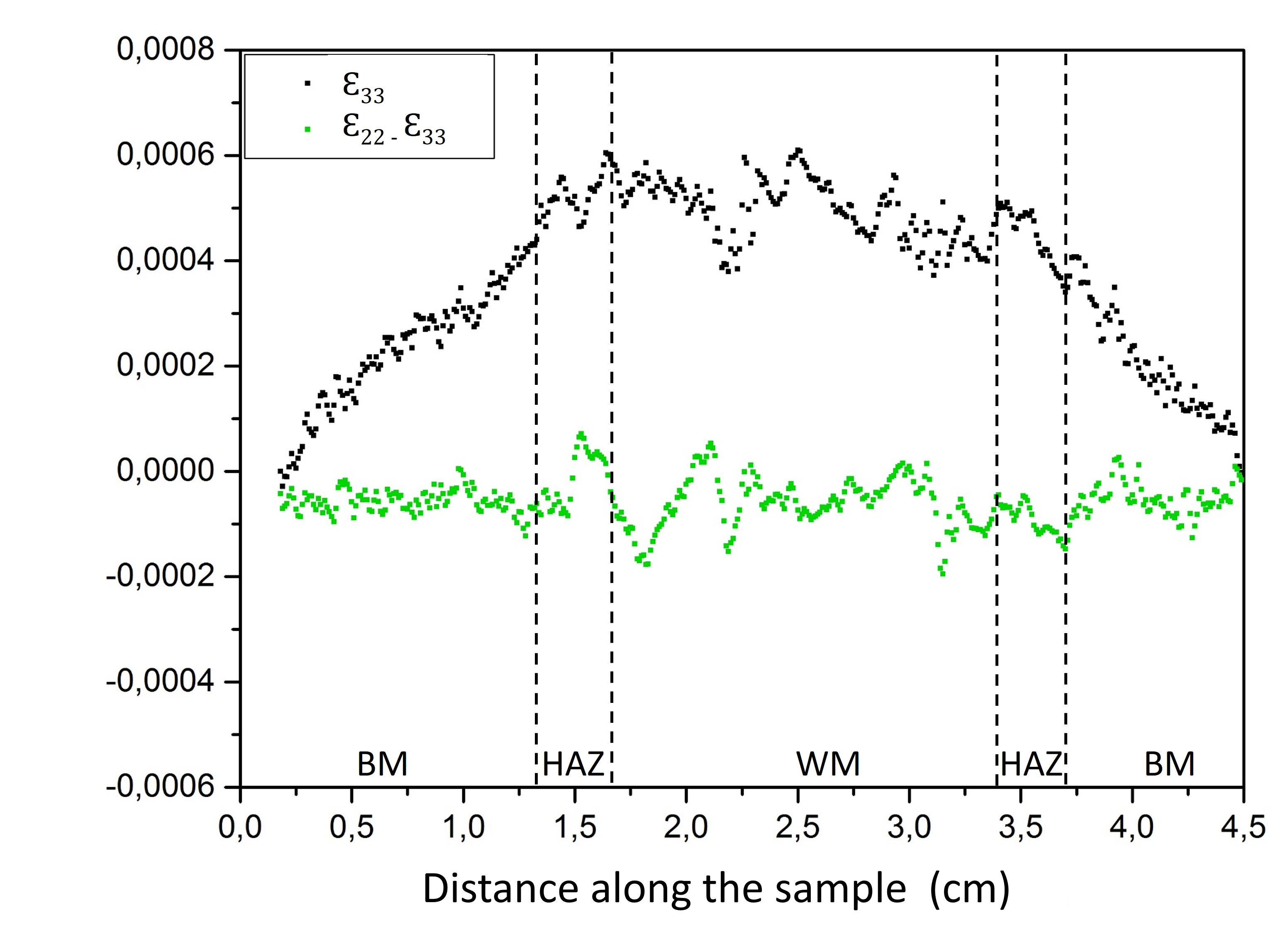}
	\caption{
		$\varepsilon_{33}$  and $\varepsilon_{22}- \varepsilon_{33}$
		evolution along the sample interrupted after 4100h creep test, 
		experimentally determined by the $\sin^{2}\Psi$ method.
	}\label{fig5}
\end{figure}

The values of $\varepsilon_{33}$  and $\varepsilon_{22}- \varepsilon_{33}$ along the welded joint are shown in Figure~\ref{fig5}. 
The $\varepsilon_{33}$ intercept exhibits an unusual variation, caused by the superposition of a temperature gradient along the sample
and a variable elastic strain. As $\varepsilon_{22}- \varepsilon_{33}$ is not sensitive to the temperature gradient, it provides more robust results. 
As it can be seen, $\varepsilon_{22}- \varepsilon_{33}$ is constant within the BM,
but exhibits strong variations within the WM and the HAZ\@. In the former case, this reveals that heterogeneities
due to the multipass welding are neither fully relieved by the post weld stress annealing nor by the creep. In the latter case,
we suppose that the variations result from internal stresses due to differences in the plastic behavior of the BM,
the HAZ and the WM\@. In the following, a particular focus will be paid on the evolution of
$\varepsilon_{22}- \varepsilon_{33}$ and $\varepsilon_{11}- \varepsilon_{33}$ along the specimens during \textit{in situ} tests. 

\begin{figure}[H]
	\centering\includegraphics[width=1\linewidth]{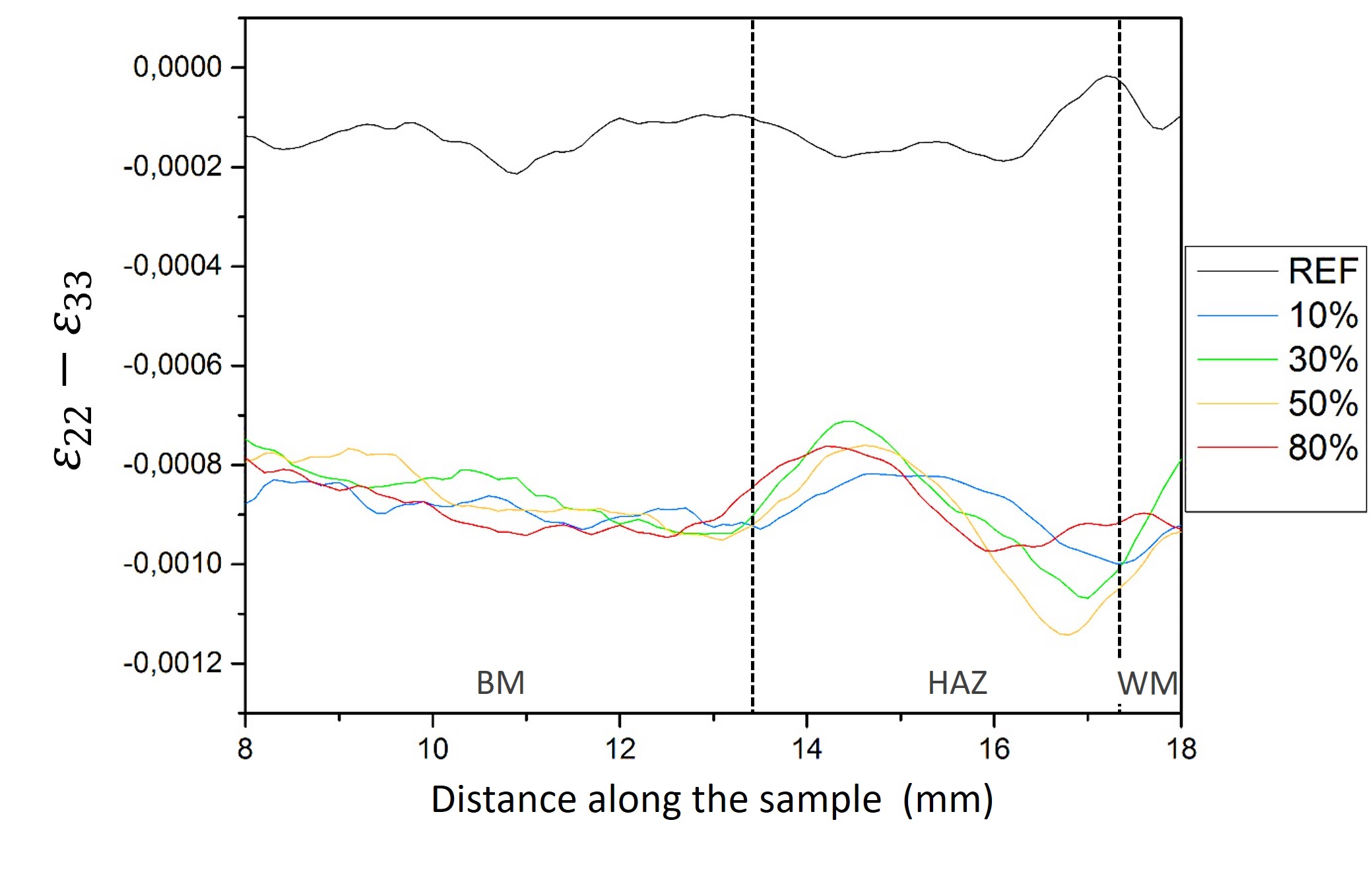}
	\caption{
		Strain evolution $\varepsilon_{22} - \varepsilon_{33}$ along the welded joint 
		after creep interrupted test.
	}\label{fig6}
\end{figure}

The evolution with time of the elastic strain $\varepsilon_{22}- \varepsilon_{33}$, along the sample (from the BM to the HAZ)
is illustrated in Figure~\ref{fig6}. The strain is averaged on 10 consecutive measurements (\textit{i.e.} 1 mm)
to reduce experimental noise. 
For the reference case, without loading, (blackline in Figure~\ref{fig6}), the strain values are shifted near 0.
Due to some dispersion on the curve $\varepsilon_{\Psi\Phi}$ as a function of $\sin^{2}\Psi$ for the reference sample,
variations in the value of the strains along the specimen are visible but are not significant. 
For all crept specimens, the strain evolves in a similar way along the welded joint,
\textit{i.e.} with a local maximum at the HAZ, bounded by local minima at the boundary with the BM
and the WM\@. Due to the repeatability of the strain evolution on 5 different samples after 5 independent experiments,
this strain evolution is significant and characteristic of creep damage on the welded joint.

\section{Modelling}\label{S:4}
\subsection{Method}
The shape of the specimens is complicated, therefore the develpment of an analytic solution is difficult
for the stress and strain fields comparison with experimental results. Thus, it is proposed to compute the plastic strain,
the total and elastic strain in the whole volume of a model specimen, then to deduce the average $\varepsilon_{22} - \varepsilon_{33}$
(resp. $\varepsilon_{11} - \varepsilon_{33}$) strain along the path of the incident X-Ray beam
travelling in the $\Phi$ = 90$^{\circ}$ (resp. $\Phi$ = 0$^{\circ}$)
direction of the $X_1X_3$ (resp. $X_2X_3$) middle plane, as shown in Figure~\ref{fig3}b. To further simplify the problem,
it is assumed that the constitutive law remains the same for the BM and the WM,
while the strain rate of the HAZ is more than double under the same conditions~\cite{spigarelli_analysis_2002}.

\subsection{FEM simulation}
\begin{figure}[H]
	\centering\includegraphics[width=1\linewidth]{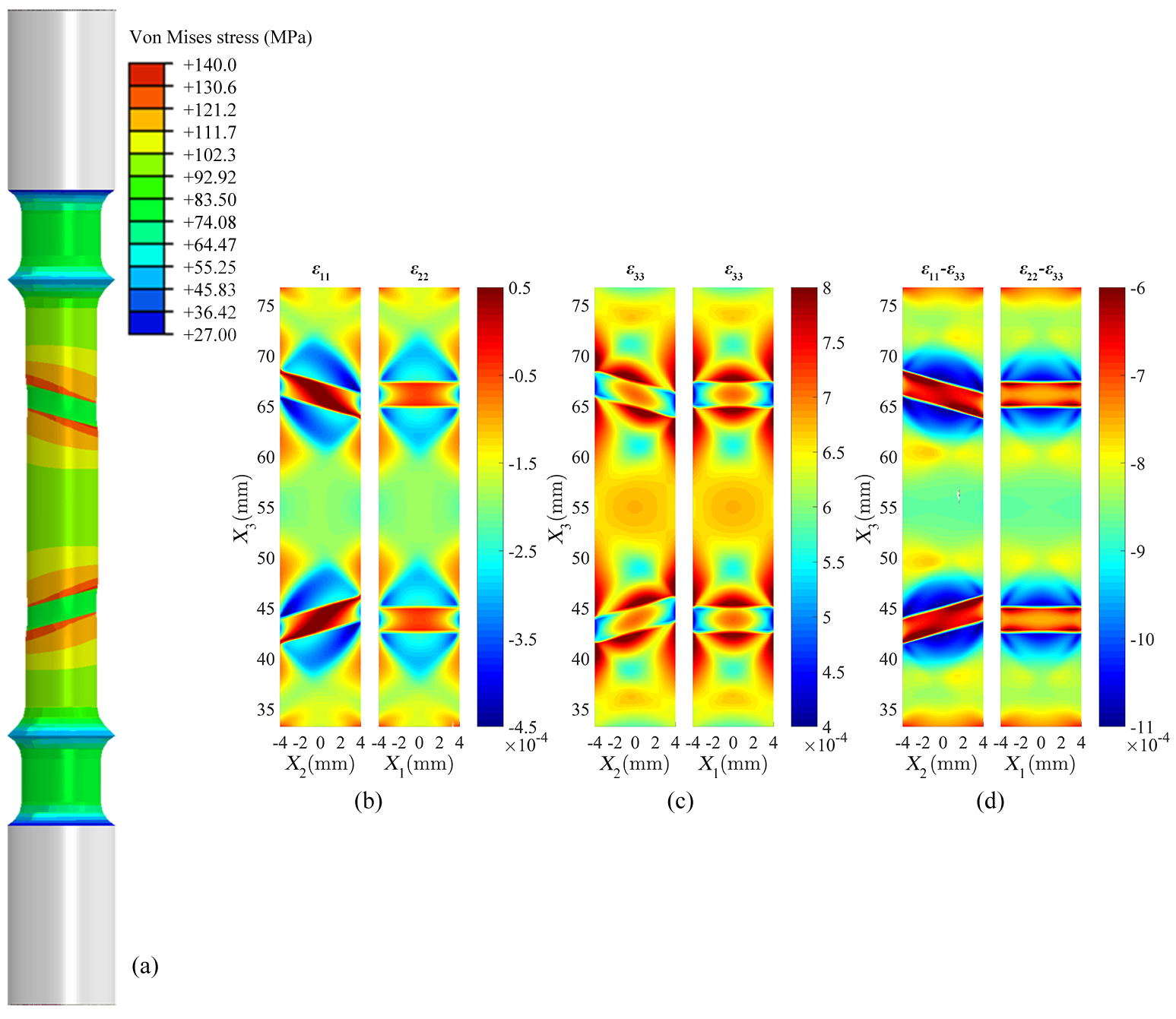}
	\caption{
		a) Von Mises stress, b)-d) calculated elastic strains in $X_1$ ($X_2$)-$X_3$ plane of the virtual sample.
	}\label{fig7}
\end{figure}

The creep tests were simulated using the Finite Elements Method with the Abaqus software package. 
The creep strain rate was taken as a power law:
\begin{equation}
\dot{\varepsilon_{eq}}=A{(\frac{\sigma_{eq}}{\sigma_0})}^{n}
\end{equation}
with $ \sigma_{eq} $ is the Von Mises stress, which is, in the case of the isotropic plasticity:  
\begin{equation}
\sigma_{eq}=\frac{1}{\sqrt{2}} \times \sqrt{{(\sigma_{1}-\sigma_{2})}^{2}+{(\sigma_{2}-\sigma_{3})}^2+{(\sigma_{3}-\sigma_{1})}^{2}}.
\end{equation}
The reference stress is $\sigma_0$=100\nbl{MPa}, and $\sigma_{1}, \sigma_{2}$ and $\sigma_{3}$ are principal stress components.
The exponent \textit{n} of the power law was fixed to $\textit{n}=10$ for all welded zones.  
Meanwhile, Young’s modulus at 600\nb$^{\circ}$C is 150\nbl{MPa}, and Poisson’s coefficient is $\nu=0.3$ for all welded zones. 
The HAZs were fixed to be 2.5 mm thick, similar to experimental specimens, with a 30$^{\circ}$ inclination versus the specimen axis. 
The parameter A was chosen so that the average strain rate is the same as the one measured during stage II off creep 
(see Figure~\ref{fig1}), for which $A_{\rm{BM}}=A_{\rm{WM}}=1.15\times10^{-6}$, and the size of the strain step between HAZ and BM 
increases with the ratio $A_{\rm{HAZ}}/A_{\rm{BM}}$. For the first step, the deformed specimen after 4100 hours of creep
is shown in Figure~\ref{fig7}{a}, with the expected (and exaggerated) necking in the vicinity of the HAZ\@. 
The (different) elastic strains in the median plane of the specimen are plotted in Figure~\ref{fig7}{b-d}. As it can be seen, 
a larger plastic strain within the HAZ results in tensile internal stresses in the $X_1$ and $X_2$ direction 
(the elastic strain is less negative than expected from the Poisson ratio only) and compressive stresses within the BM
and the WM\@. The magnitude of the variations of $\varepsilon_{33}$ is lower, 
as these variations result mainly from the limited area reduction at necking and from compatibility stresses at surfaces. 
Last, the $\varepsilon_{22}- \varepsilon_{33}$ and $\varepsilon_{11} - \varepsilon_{33}$ plots exhibit a stronger contrast.

\begin{figure}[H]
	\centering\includegraphics[width=1\linewidth]{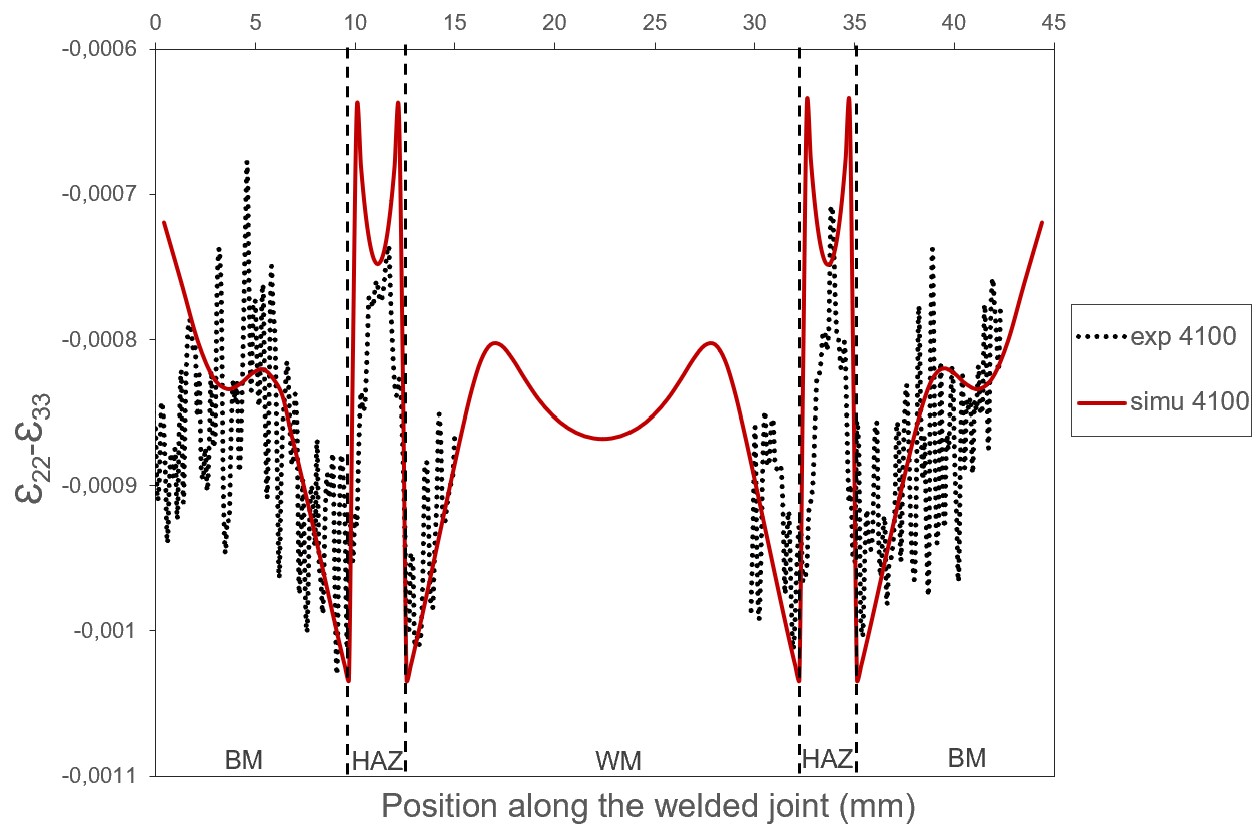}
	\caption{
		Experimental (dotted) and calculated (plain line)
		$\varepsilon_{22}- \varepsilon_{33}$ strain curves after 4100 hours.
	}\label{fig8}
\end{figure}

The variation of the slopes  $\varepsilon_{22}- \varepsilon_{33}$  along the specimen axis at t = 4100 hours
between experimental and simulated values, for the parameters chosen in the previous paragraph, are shown in Figure~\ref{fig8}.
The experimental curve is derived from the raw data, which explains the high level of noise.
The experimental curve of the WM is not plotted, due to the large number of small grains and 
the presence of gas holes in this area, which increase noise and random variations.
In Figure~\ref{fig8}, simulated and experimental curves fit with a good representability of the peak in HAZ\@. 

\begin{figure}[H]
	\centering\includegraphics[width=1\linewidth]{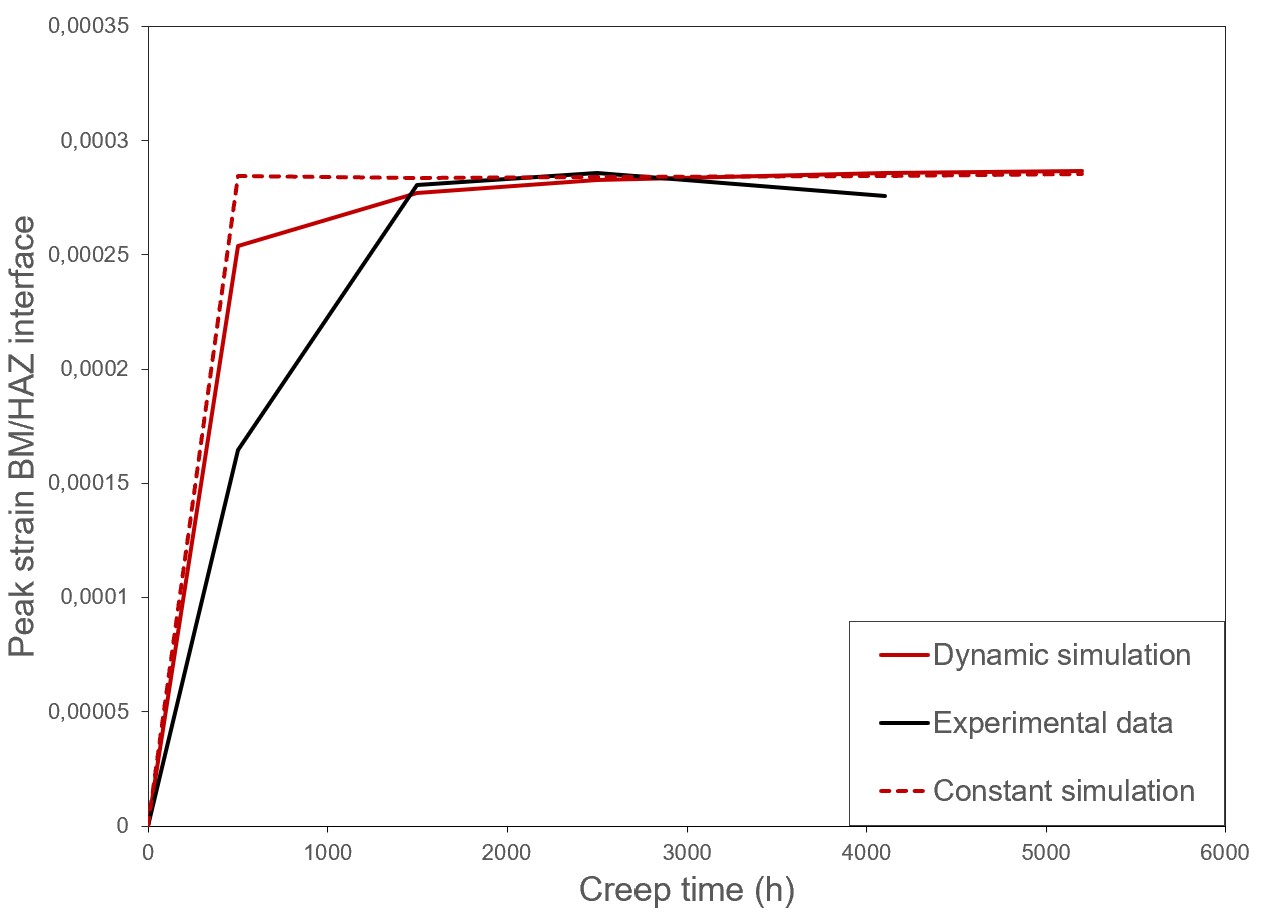}
	\caption{
		Step of $\varepsilon_{22}- \varepsilon_{33}$ at the BM/HAZ interface
		as a function of creep time for experimental and simulation data. Simulation curves are performed
		for dynamic (plain line) and constant (dashed line) strain rate.
	}\label{fig9}
\end{figure}

The evolution of the step, at the BM/HAZ interface as a function of creep time,
illustrates the damage kinetics of this zone. This evolution is illustrated in Figure~\ref{fig9}, 
with the comparison of the simulated (red and dashed red lines) and experimental (black line) values. 
Experimentally, the step at the BM/HAZ interface is visible from the end of creep stage I,
and increases up to 2500 hours, \textit{i.e.} in the middle of stage II, and then saturates. In the simulation,
the initial HAZ strain rate is chosen to be constant as $A_{\rm{HAZ}}=A_{\rm{BM}} \times 200$,
resulting in stabilisation within a few hours as shown in Figure~\ref{fig9} by the red dotted line.
To obtain a better fit, the parameter  $ A_{\rm{HAZ}}$ was allowed to vary with time in order to simulate a gradual softening:
\begin{equation}
A_{\rm{HAZ}} (t)=A_{\rm{BM}}+A_{\rm{HAZ}} (1-\exp(-t/t_{0} ))
\end{equation}

The time constant $ t_{0} $ was taken as 1000 hours and  $ A_{\rm{HAZ}} = A_{\rm{BM}}~\times~200$. With this gradual softening,
the simulated deformation kinetic fits more closely the experimental one. 

\section{Discussion}\label{S:5}

Strain along the weld evolves with a local maximum at the HAZ, 
bounded by local minima at the interface with the  BM and the WM\@. 
These experimental results were confirmed by FEM simulations
using a power law for the creep strain rate. 
However, to match the experimental results, the strain rate of the HAZ 
must be larger than that of the BM and also variable to follow the kinetics. 

The strain evolution is driven by a microstructural softening, which is not constant 
over the time of creep. 
The nucleation of initial cavities is due to strong local plasticity related to the softening of the material. 
Nevertheless, according to the SEM analysis (Figure~\ref{fig4}), the cavities are located in the 
zone with the highest deformation, \textit{i.e.} the ICHAZ, 
but exclusively in the core of the sample. This phenomenon is also found through FEM simulations as shown in Figure~\ref{fig7}, the HAZ deforms more than the others, resulting in internal stresses, and the distribution of creep strain on the HAZ varies from the surface toward the center.

\begin{figure}[H]
	\centering\includegraphics[width=1\linewidth]{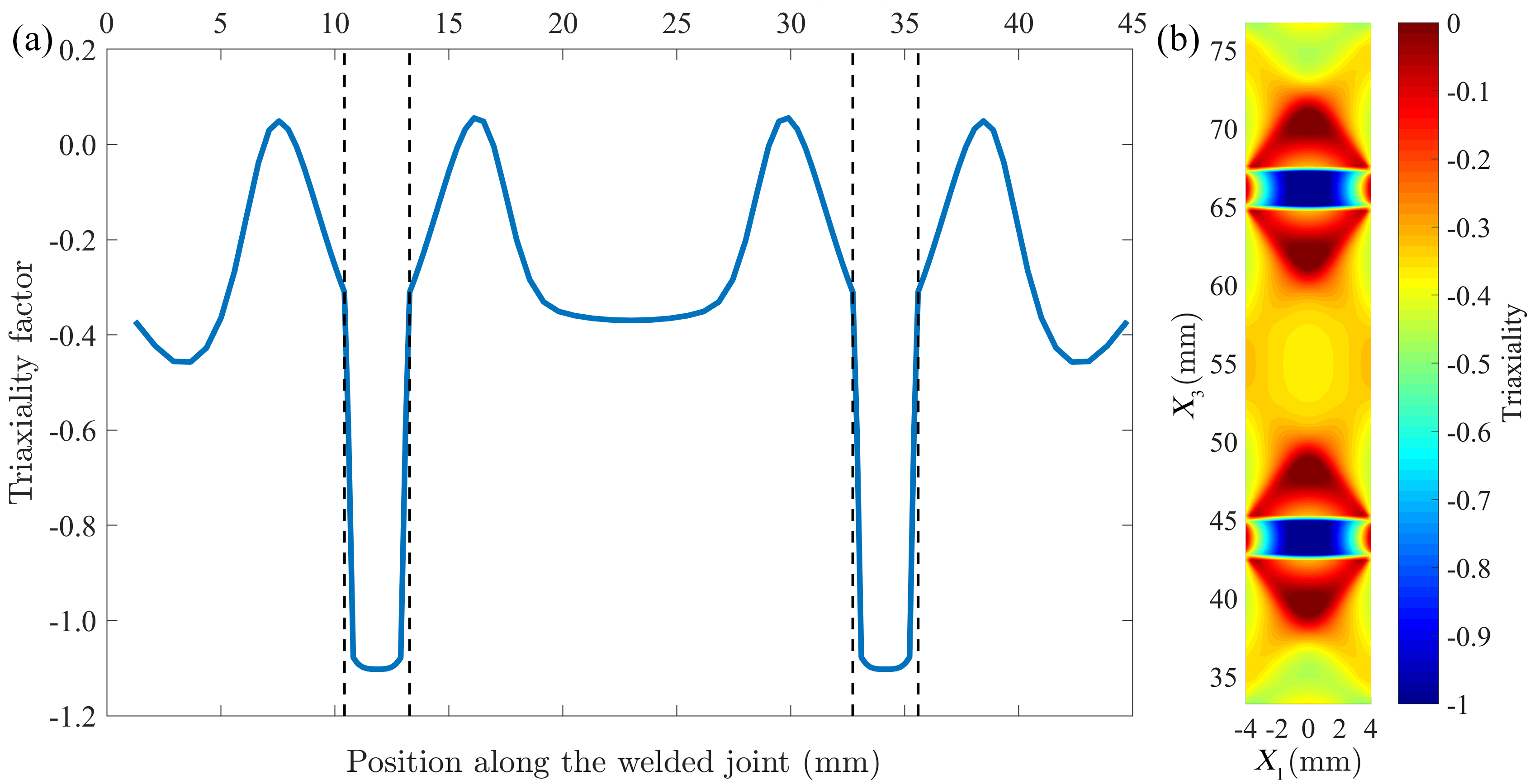}
	\caption{Simulated TF as a function of position along the welded joint.}\label{fig10}
\end{figure}

Some authors found that the concentration of the deformations and the high triaxiality rate 
affect the distribution of voids~\cite{abd_el-azim_long_2013, watanabe_creep_2006, li_evaluation_2009,  ogata_damage_2010}. The triaxiality factor (TF) is the ratio between {the} hydrostatic stress: 
\begin{equation}
    \sigma_m=(\sigma_{11}+\sigma_{22}+\sigma_{33})/3
\end{equation}
and {the} Von Mises stress $\sigma_{eq}$. TF is thus defined by~\cite{ogata_damage_2010}: 
\begin{equation}
    \rm{TF}=\sigma_m/\sigma_{eq} 
\end{equation}
The distribution of the TF along the welded joint 
after 4100h creep is shown in Figure~\ref{fig10}. 
The sign of the TF depends on the sign of the hydrostatic stress. 
In our case the TF is positive in tension and negative in compression. 
TF is lowest in the HAZ, with a jump {on} the interfaces with other zones.
A large TF favors brittle fracture by cleavage~\cite{triaxiality_damage_initiation} which was experimentally observed in Figure~\ref{fig4},
where cavities and undeformed grains are present. HAZs exhibit maximum creep strain and minimum of the TF,
favoring creep damage with cavities in the core of the welded joint.

\section{Conclusion}\label{S:6}
The \textit{in situ} tests combined with High Energy X-Ray Diffraction, using synchrotron radiation,
allow to determine internal deformations.
The strain along the welded joint evolves with a local maximum at the HAZ,
{limited} by local minima at the boundary with the BM and the WM\@.
The difference in strain rates between the HAZ and other welded joint {zones}
{results in} maximum triaxial stresses in this area. 
These triaxial stresses are responsible for cavity nucleation and growth in the core of the ICHAZ,
further reducing the mechanical strength of the ICHAZ\@.
The combination of cavity formation and recovery of the microstructure~\cite{collomb_characterization_nodate}
{leads to severe local softening} in the ICHAZ,
corresponding to the failure zone.

\section*{Acknowledgment}
This work has been sponsored by Institut de Soudure through its research center in Yutz France and l’Institut Carnot ICÉEL\@.
The authors thank Dr.\ Stéphane Berbenni and Dr.\ Thiebaud Richeton (LEM3) for fruitful discussions.
%



\bibliographystyle{unsrt}
\bibliography{Coll_Chen_arxiv.bib}






\end{document}